\RequirePackage{fix-cm}
\documentclass[twocolumn, epjc3]{svjour3}
\usepackage[square, numbers, comma, sort&compress]{natbib}

\journalname{Eur. Phys. J. C}

\usepackage{latexsym}
\usepackage{amsmath}
\usepackage{amssymb}
\usepackage{amsfonts}
\usepackage{CJKutf8}

\usepackage[mathscr,scaled=1.15]{urwchancal}
\DeclareFontFamily{OT1}{pzc}{}
\DeclareFontShape{OT1}{pzc}{m}{it}%
{<-> s * [1.15] pzcmi7t}{}
\DeclareMathAlphabet{\mathpzc}{OT1}{pzc}{m}{it}

\usepackage{color}

\usepackage{supertabular}
\usepackage{placeins}
\usepackage{epsfig}
\usepackage{graphicx}

\definecolor{purple}{rgb}{0.5,0,0.5}
\definecolor{blue}{rgb}{0.0,0,0.9}
\definecolor{prdblue}{rgb}{0.133,0.118,0.498}
\usepackage[colorlinks=true, pdfstartview=FitV, linkcolor=prdblue, citecolor= prdblue, urlcolor=prdblue]{hyperref}

\begin{document}


\begin{CJK}{UTF8}{song}

\title{$\,$\\[-7ex]\hspace*{\fill}{\normalsize{\sf\emph{Preprint no}. NJU-INP 044/21}}\\[1ex]
Vector-meson production and vector meson dominance}

\author{
        Y.-Z.\ Xu\thanksref{eYZXu,NJU,INP} 
        \and
        S.-Y.\ Chen\thanksref{eSYChen,NJU,INP} 
        \and
       Z.-Q.\ Yao\thanksref{eZQYao,NJU,INP} 
        \and \\
        D.\ Binosi\thanksref{eDB,ECT}
    \and
       Z.-F.\ Cui\thanksref{eZFC,NJU,INP} 
       \and
       C.\,D.\ Roberts\thanksref{eCDR,NJU,INP}
}

\thankstext{eYZXu}{xuyz@smail.nju.edu.cn}
\thankstext{eSYChen}{siyangchen@smail.nju.edu.cn}
\thankstext{eZQYao}{zqyao@smail.nju.edu.cn}
\thankstext{eDB}{binosi@ectstar.eu}
\thankstext{eZFC}{phycui@nju.edu.cn}
\thankstext{eCDR}{cdroberts@nju.edu.cn}

\authorrunning{Yin-Zhen Xu \emph{et al}.} 

\institute{School of Physics, Nanjing University, Nanjing, Jiangsu 210093, China \label{NJU}
           \and
           Institute for Nonperturbative Physics, Nanjing University, Nanjing, Jiangsu 210093, China \label{INP}
           \and
           European Centre for Theoretical Studies in Nuclear Physics
and Related Areas,\\
\hspace*{1.1ex}Villa Tambosi, Strada delle Tabarelle 286, I-38123 Villazzano (TN), Italy\label{ECT}
            }

\date{2021 July 07}

\maketitle
\end{CJK}

\begin{abstract}
We consider the fidelity of the vector meson dominance (VMD) assumption as an instrument for relating the electromagnetic vector-meson production reaction $e + p \to e^\prime + V + p$ to the purely hadronic process $V + p \to V+p$.  Analyses of the photon vacuum polarisation and the photon-quark vertex reveal that such a VMD \emph{Ansatz} might be reasonable for light vector-mesons.  However, when the vector-mesons are described by momentum-dependent bound-state amplitudes, VMD fails for heavy vector-mesons: it cannot be used reliably to estimate either a photon-to-vector-meson transition strength or the momentum dependence of those integrands that would arise in calculations of the different reaction amplitudes.  Consequently, for processes involving heavy mesons, the veracity of both cross-section estimates and conclusions based on the VMD assumption should be reviewed, \emph{e.g}., those relating to hidden-charm pentaquark production and the origin of the proton mass.
 \end{abstract}


\section{Introduction}
\label{Introduction}
The interaction of a heavy vector-meson, $J/\psi$ or $\Upsilon$, with a proton target offers prospects for access to a QCD van der Waals interaction, generated by multiple gluon exchange \cite{Brodsky:1989jd, TarrusCastella:2018php}, and the QCD trace anomaly \cite{Luke:1992tm, Kharzeev:1995ij}.  The former is of interest because it may relate to, \emph{inter alia}, the observation of hidden-charm pentaquark states \cite{Aaij:2015tga}; whereas the latter has received attention owing to its connection with emergent hadron mass (EHM), the phenomenon that is seemingly responsible for roughly 99\% of the visible mass in the Universe \cite{Aguilar:2019teb, Chen:2020ijn, Krein:2020yor, Arrington:2021biu, Roberts:2021nhw}.

In the absence of vector-meson beams, experiments at modern electron ($e$) accelerators attempt to access such interactions via electromagnetic production of vec-tor-mesons ($V$) from the proton ($p$), in reactions like $e + p \to e^\prime + V + p$ \cite{Ali:2019lzf}; and the same method is proposed for use at planned higher-energy facilities \cite{Anderle:2021wcy, AbdulKhalek:2021gbh}.  In this connection, it is typically assumed that single-pole vector meson dominance (VMD) \cite{Sakurai:1960ju, sakurai:1969, Fraas:1970vj} may reliably be used to draw a direct link between the electromagnetic production process and the $V p\to V p$ cross-section.
Namely, as illustrated in Fig.\,\ref{FigVMD}, the interaction is supposed to take place in a sequence of steps:
\begin{figure}[h] 
\centerline{\includegraphics[clip,width=0.4\textwidth]{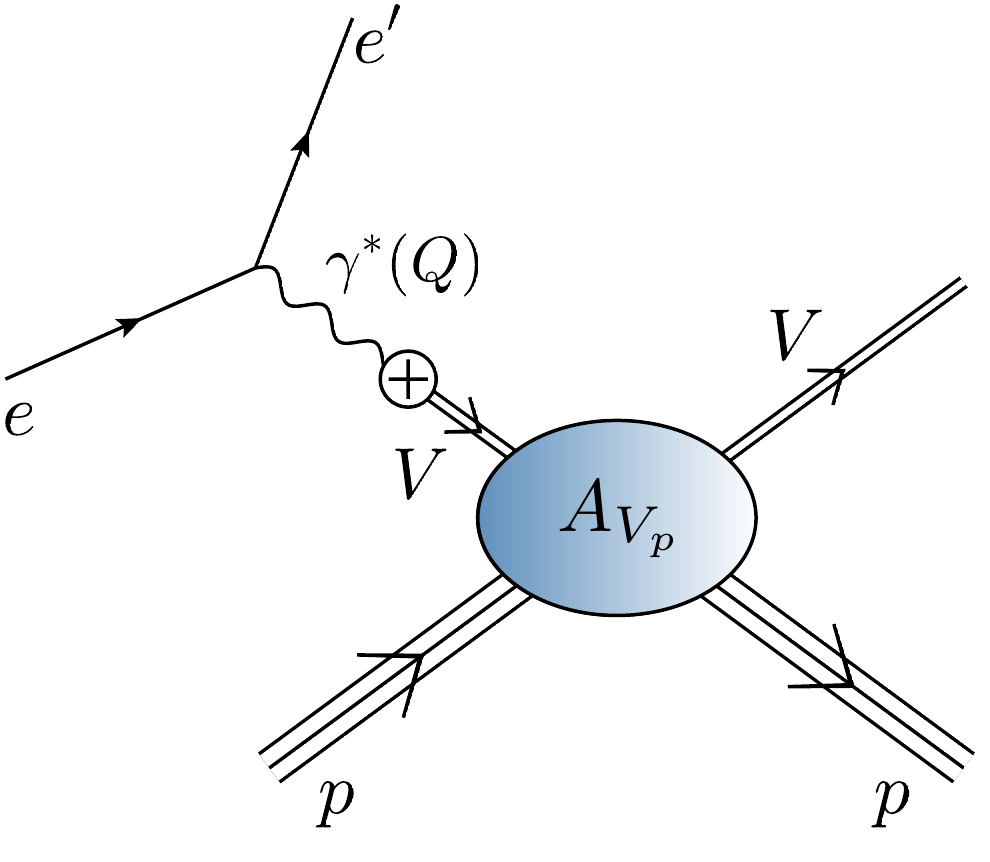}}
\caption{\label{FigVMD} Electromagnetic production of a vector-meson from the proton: $e + p \to e^\prime + V + p$, interpreted as providing access to $V + p \to V+p$ using a vector meson dominance model.  The VMD transition $\gamma^{(\ast)}(Q)\to V$ is indicated by the crossed-circle and is typically assumed to occur with a momentum-independent strength $\gamma_{\gamma V}$ \cite{Sakurai:1960ju, sakurai:1969, Fraas:1970vj}.
%
}
\end{figure}
\linebreak (\emph{a}) $e \to e^\prime + \gamma^{(\ast)}(Q)$;
(\emph{b}) $\gamma^{(\ast)}(Q)\to V$;
and (\emph{c}) $V+ p \to V+p$.  Step (\emph{b}) expresses the VMD hypothesis.
As commonly used, it assumes: (\emph{i}) that a photon, which is, at best, real, but is generally spacelike, so that $Q^2 \geq 0$, transmutes into an on-shell vector-meson, with timelike momentum $Q^2=-m_V^2$;
and (\emph{ii}) that the $Q^2\geq 0$ strength and character of the transition in (\emph{b}) is unchanged from that measured in the real process of vector-meson decay, $V \to \gamma^\ast(Q^2=-m_V^2) \to e^+ + e^- $, \emph{i.e}., $\gamma_{\gamma V}$ is supposed to remain fixed at its meson on-shell value and acquire no momentum dependence.
(Throughout, we use the Euclidean metric conventions specified in Ref.\,\cite[Sec.\,2.1]{Roberts:2000aa}.)

The VMD \emph{Ansatz} was introduced before the development of quantum chromodynamics (QCD) for use in analysing energetic electromagnetic interactions of light hadrons, \emph{viz}.\ states with masses not much different from that of the proton \cite{Sakurai:1960ju, sakurai:1969, Fraas:1970vj}.  It is still used today and for a much wider range of systems \cite{Wu:2019adv, Gryniuk:2020mlh} because the alternative is to develop a sophisticated, nonperturbative reaction theory that can explain quark+antiquark scattering from hadron targets into vector-meson final-states.  That nettle has not yet been grasped.  Notwithstanding this, the lack of an alternative does not validate the VMD expedient; and its fidelity should be reconsidered with QCD constraints in mind.

In Sec.\,\ref{SecPVP}, we review connections between the dilepton decay of vector mesons and the quark contribution to the photon vacuum polarisation, arriving via this route at a result first highlighted in Ref.\,\cite{Feldman:1963fxb}.  Namely, straightforward implementation of any photon--vector-meson current-field identity \cite{Sakurai:1960ju, sakurai:1969} must necessarily generate a tachyonic photon; hence, cannot alone be used to develop or support a VMD phenomenology.
With that route to VMD closed, Sec.\,\ref{SecGQvtx} turns to a discussion of the photon-quark vertex, $\Gamma_\nu^\gamma(k;Q)$, highlighting that all vector-mesons appear as a pole in this vertex at a timelike value of the total momentum: $Q^2 = -m_V^2$, where $m_V$ is the meson mass.  Then the persistence of this link between the photon and vector-meson to $Q^2=0$, away from the meson mass-shell, is explored using two \emph{Ans\"atze} for the quark+antiquark scattering matrix.
Section~\ref{epilogue} presents a summary and perspective.

\section{Photon vacuum polarisation}
\label{SecPVP}
The physical process $V\to e^+ e^-$ is described by a leptonic decay constant, $f_V$, which can be expressed as follows \cite{Ivanov:1998ms, Maris:1999nt}:
\begin{align}
f_V m_V \epsilon_\mu^\lambda(Q) & = {\rm tr}_{\rm CD} Z_2 \int_{dk}^\Lambda\,
\gamma_\mu \chi_V^\lambda(k,Q)\,,
\label{EqfVmV}
\end{align}
where the trace is over colour and spinor indices;
$\epsilon_\mu^\lambda(Q)$ is the vector-meson polarisation vector,
\begin{equation}
\label{polarisation}
\sum_{\lambda =-1,0,1} \epsilon_\mu^\lambda(Q)\epsilon_\nu^\lambda(Q)
= \delta_{\mu\nu} - Q_\mu Q_\nu/Q^2=:T_{\mu\nu}(Q)\,,
\end{equation}
with $Q^2= -m_V^2$ for the on-shell meson;
$\int_{dk}^\Lambda$ represents a translationally-invariant regularisation of the four-dimensional integral, with $\Lambda$ the regularisation scale;
$Z_2(\zeta,\Lambda)$ is the quark wave function renormalisation constant, with $\zeta$ the renormalisation scale for all QCD quantities considered herein;
and the meson's Poincar\'e-covariant Bethe-Salpeter wave function is
\begin{equation}
\chi_V^\lambda(k,Q) = S(k_+) \Gamma_V^\lambda(k;Q) S(k_-)\,,
\end{equation}
where $S(k)$ is the dressed-propagator for the valence-quark/-antiquark from which $V$ is constituted, $k_+ = k+\eta Q$, $k_- = k -(1-\eta) Q$, $0\leq \eta\leq 1$, and $\Gamma_V^\lambda(k;Q) $ is the vector-meson's canonically-normalised Bethe-Salpeter \linebreak amplitude \cite{LlewellynSmith:1969az, Nakanishi:1969ph}.  (In our normalisation, the pion's leptonic decay constant is $f_\pi \approx 0.092\,$GeV; and we renormalise at $\zeta=19\,$GeV, using a scheme that is independent of current-quark mass \cite{Chang:2008ec}.)

In terms of its decay constant, a vector-meson's $e^+e^-$ decay width is
\begin{equation}
\Gamma_{V\to e^+ e^-} = \frac{8\pi\alpha_{\rm em}^2}{3} \frac{f_V^2}{m_V} \bar e_V^2 \,,
\end{equation}
where $\alpha_{\rm em}=e^2/(4\pi)$ is the fine structure constant of quantum electrodynamics (QED) and $\bar e_V^2$ is a squared sum of quark charges weighted by the meson's flavour wave function, as referred to the positron charge:
\begin{equation}
2 (\bar e_{\rho^0}^2, \bar e_\omega^2, \bar e_\phi^2, \bar e_{J/\psi}^2, \bar e_\Upsilon^2)
= (1, \tfrac{1}{9}, \tfrac{2}{9}, \tfrac{8}{9}, \tfrac{2}{9} )\,.
\end{equation}
These values assume isospin symmetry and ideal mixing for vector-mesons, \emph{e.g}., the $\phi$-meson is a $s\bar s$ system.

A dimensionless coupling for vector-mesons is also commonly used:
\begin{equation}
g_V = \frac{m_V}{f_V}\,.
\end{equation}
The $ \surd 2\bar e_V$ factor is sometimes absorbed into $1/g_V$; and the current-field identity that is definitive of VMD is expressed via
\begin{equation}
\label{EqCFI}
\gamma_{\gamma V} = e    \frac{m_V^2}{g_V}\,.
\end{equation}

The decay constants, $f_V$, control the strength of any photon--vector-meson mixing.  This can be seen by considering the associated quark-loop contribution to the photon vacuum polarisation tensor, the regularised expression for which is
\begin{align}
\Pi^\prime_{\mu\nu}(Q) & = \bar e_V^2 {\rm tr}_{\rm CD} Z_2
\int_{dk}^\Lambda \gamma_\mu S(k_+)\Gamma_\nu^\gamma(k,Q)S(k_-) \,,
\label{EqPiQ}
\end{align}
where the photon-quark vertex satisfies the following Ward-Green-Takahashi identities:
\begin{subequations}
\label{SWGTI}
\begin{align}
i \Gamma_\nu^\gamma(k,Q=0) & = \partial_\nu S^{-1}(k)\,, \label{Ward}\\
Q_\nu i\Gamma_\nu^\gamma(k,Q) & = S^{-1}(k_+) - S^{-1}(k_-)\,. \label{WGTI}
\end{align}
\end{subequations}
Given the general form of the quark propagator,
\begin{equation}
\label{GenS}
S(k) =\frac{1}{i\gamma\cdot k A(k^2)+ B(k^2)} \equiv \frac{Z(k^2)}{i\gamma\cdot k + M(k^2)}\,,
\end{equation}
where $M(k^2)$ is renormalisation point invariant, Eq.\,\eqref{Ward} entails that the $Q^2\simeq 0$ photon-quark vertex is completely described by three or less tensor structures and associated scalar functions, all of which are fully determined by those appearing in the quark propagator \cite{Ball:1980ay, Curtis:1990zs}.  This hints that any overlap between the $Q^2=0$ quark-photon vertex and the vector-meson bound-state is both indirect and modest at best \cite{Roberts:2000aa, Maris:1999bh}.

Using Eq.\,\eqref{WGTI} and capitalising on the properties of $\int_{dk}^\Lambda$, one may readily establish Ward-Green-Takahashi identities for the photon vacuum polarisation:
\begin{equation}
\label{EqPiWGTI}
Q_\mu \Pi_{\mu\nu}^\prime(Q) = 0 = \Pi_{\mu\nu}^\prime (Q) Q_\nu \,,
\end{equation}
which entail
\begin{equation}
\Pi_{\mu\nu}^\prime(Q) = T_{\mu\nu}(Q) \, Q^2 \, \Pi^\prime(Q^2)\,.
\end{equation}

Following convention, the renormalised photon vacuum polarisation tensor is
\begin{subequations}
\label{QEDVP}
\begin{align}
\Pi_{\mu\nu}(Q) & = T_{\mu\nu}(Q) \, Q^2\,\Pi(Q^2) \,,\\
\Pi(Q^2) & =  \Pi^\prime(Q^2) - \Pi^\prime(Q^2=0)
\end{align}
\end{subequations}
so that $\Pi(Q^2) =0$ and the photon remains massless in the renormalised theory.  The absence of a mass-scale in the photon polarisation tensor is an empirical fact.

The photon vacuum polarisation is connected to vector-mesons through the dressed photon-quark vertex.  In the neighbourhood of a vector-meson pole, ignoring any hadronic width:
\begin{align}
\frac{1}{Z_2}&[\Gamma_\nu^\gamma(k,Q)]_{tu} \stackrel{Q^2 + m_V^2 \simeq 0}{=} \mbox{regular~terms} \nonumber  \\
&
 +
\int_{dl}^\Lambda
\, 2\frac{[\Gamma_V^\lambda(k;Q)]_{ut} [\Gamma_V^\lambda(l;Q)]_{rs}}{Q^2+m_V^2}
[S(l_-) \gamma_\nu S(l_+) ]_{sr}\,, \label{EqVMDvertex}
\end{align}
where we have exploited properties of the appearing objects under charge conjugation.  (See, e.g., Ref.\,\cite[Appendix~A]{Bhagwat:2007rj}.)  Inserting this result into Eq.\,\eqref{EqPiQ} and recognising that the regularising term receives no bound-state contribution, one finds
\begin{subequations}
\begin{align}
&\Pi_{\mu\nu}(Q)  \stackrel{Q^2 + m_V^2 \simeq 0}{=} \nonumber \\
& \quad \bar e_V^2 \,
{\rm tr}_{\rm CD} Z_2
\int_{dk}^\Lambda \gamma_\mu S(k_+)\Gamma_V^\lambda(k;Q) S(k_-) \nonumber \\
& \times \frac{2}{Q^2+m_V^2}
{\rm tr}_{\rm CD} Z_2
\int_{dl}^\Lambda \gamma_\nu S(l_+)\Gamma_V^\lambda(k;Q)  S(l_-) \\
& =\bar e_V^2  \frac{2 f_V^2 m_V^2}{Q^2+m_V^2} T_{\mu\nu}(Q)\,,
\end{align}
\end{subequations}
where Eq.\,\eqref{polarisation} has been used.  Consequently,
\begin{equation}
Q^2 \, \Pi(Q^2) \stackrel{Q^2 + m_V^2 \simeq 0}{=}
\bar e_V^2\, \frac{2 f_V^2 m_V^2}{Q^2+m_V^2}\, ;
\end{equation}
and, therefore, on $Q^2+m_V^2 \simeq 0$, the timelike photon is indistinguishable from the vector-meson.  This is merely the statement that $e^+ e^-$ collisions with a tuned centre-of-mass energy can be used to produce vector-mesons.

However, the VMD assumption asserts  that the photon is also indistinguishable from the vector-meson on $Q^2\simeq 0$.  This is the content of Eq.\,\eqref{EqCFI} and it is plainly false because $\left.Q^2 \Pi(Q^2) \right|_{Q^2\simeq 0}\equiv 0 $ as a consequence of photon masslessness: a massive composite vector-meson cannot be confused (mix) with an on-shell massless photon.  With this contradiction, we recover the objection to Eq.\,\eqref{EqCFI} that was first raised in Ref.\,\cite{Feldman:1963fxb}.   A remedy for this flaw was proposed in Ref.\,\cite{Kroll:1967it}.  It consists in constructing a local Lagrangian for pointlike photons, vector-mesons, and nucleons, with couplings tuned to ensure cancellation of the photon mass term that is generated by the interaction in Eq.\,\eqref{EqCFI}.  In the context of QCD, whose interactions do not generate pointlike hadrons, this solution is untenable.  We therefore ask: is there a QCD alternative?

\section{Photon+quark vertex}
\label{SecGQvtx}
Another place to look for justification of the VMD assumption is suggested by Eq.\,\eqref{EqVMDvertex}.  The dressed photon-quark vertex describes precisely how a photon (timelike, real, or spacelike) couples to a quark+anti\-quark pair; and this is the general character of the interaction expressed in Fig.\,\ref{FigVMD}:
(\emph{a}$^\prime$) $e \to e^\prime + \gamma^\ast(Q)$;
(\emph{b}$^\prime$) $\gamma^\ast(Q) \to q + \bar q$;
and (\emph{c}$^\prime$) $q + \bar q + p \to V + p$.  Plainly, $\Gamma_\nu^\gamma(k;Q)$ is momentum dependent.
So, the question to be addressed is: are there any conditions under which $\left.\Gamma_\nu^\gamma(k;Q)\right|_{Q^2\simeq 0}$ has a link to an on-shell vector-meson that may be approximated by Eq.\,\eqref{EqCFI} or something similar?  To answer this, one must compute $\Gamma_\nu^\gamma(k;Q)$.

The contribution of a given quark flavour to the dressed-quark photon vertex can be obtained by solving the following integral equation:
\begin{align}
[\Gamma_\nu^\gamma &(k;Q)]_{tu}  = Z_2 [\gamma_\nu]_{tu}  \nonumber \\
& +
\int_{dl}^\Lambda {\mathpzc K}_{tu}^{rs}(k,l;P)
[S(l_+) \Gamma_\nu^\gamma(l;Q)S(l_-)]_{sr}, \label{EqBS}
\end{align}
where ${\mathpzc K}(k,l;P) $ is the quark+antiquark scattering kernel.  After approximately thirty years of study, much has been learnt about this interaction \cite{Binosi:2014aea, Binosi:2016wcx, Cyrol:2017ewj, Cui:2019dwv, Qin:2020jig}; and we subsequently discuss solutions of Eq.\,\eqref{EqBS} and the associated, coupled quark gap equation obtained using two physically motivated choices.

Our calculations are performed using rainbow-ladder (RL) truncation \cite{Munczek:1994zz, Bender:1996bb}, which is the leading-order in the most widely used approximation scheme for QCD's Dyson-Schwinger equations (DSEs) \cite{Eichmann:2016yit, Fischer:2018sdj, Roberts:2020udq, Eichmann:2020oqt, Roberts:2020hiw, Qin:2020rad}.  It is known to deliver realistic predictions for, \emph{inter alia}, the properties of ground-state vector-mesons constituted from degenerate valence de\-grees-of-freedom for reasons that are well understood \cite{Eichmann:2016yit, Qin:2020rad, Qin:2020jig}.  In RL truncation, the quark + antiquark scattering kernel can be written ($q = k-l$):
\begin{subequations}
\label{KDinteraction}
\begin{align}
{\mathpzc K}_{tu}^{rs}(k,l;P)
& = {\mathpzc G}_{\mu\nu}(q) [i\gamma_\mu]_{ts} [i\gamma_\nu]_{ru}\,,\\
 {\mathpzc G}_{\mu\nu}(q)  & = \tilde{\mathpzc G}(q^2) T_{\mu\nu}(q)\,.
\end{align}
\end{subequations}
The form of $\tilde{\mathpzc G}(q^2)$ is the key to drawing connections with QCD and all reasonable \emph{Ans\"atze} express features deriving from its non-Abelian character.

\subsection{Contact interaction}
\label{SecCI}
Owing to the emergence of a gluon mass-scale in QCD, enabled by strong non-Abelian gauge-sector dynamics \cite{Boucaud:2011ug, Aguilar:2015bud, Binosi:2016xxu, Gao:2017uox, Cui:2019dwv, Huber:2018ned}, $\tilde{\mathpzc G}$ is nonzero and finite at infrared momenta; so, one may write
\begin{align}
\label{GSCI}
\tilde{\mathpzc G}(q^2) & \stackrel{k^2 \simeq 0}{=} \frac{4\pi \alpha_{\rm IR}}{m_G^2}\,.
\end{align}
QCD has \cite{Cui:2019dwv}: $m_G \approx 0.5\,$GeV, $\alpha_{\rm IR} \approx \pi$.

Translating the model of Ref.\,\cite{Nambu:1961tp} into a Schwinger function framework typical of modern continuum approaches to QCD, Eqs.\,\eqref{KDinteraction}, \eqref{GSCI} have been used as the starting point for development of a symmetry-preserving formulation of a vector$\times$vector contact interaction (SCI) \cite{Roberts:2010rn, Roberts:2011wy}.  This kernel maintains the character of more realistic treatments of the continuum bound-state problem whilst, nevertheless, introducing an algebraic simplicity.  The many applications can be traced from \linebreak Refs.\,\cite{Zhang:2020ecj, Yin:2021uom, Lu:2021sgg, Xu:2021iwv}.

\begin{table}[t]
\caption{\label{Tab:DressedQuarks}
The SCI is defined by three parameters.  The infrared cutoff (confinement scale) is fixed: $\Lambda_{\rm ir} = 0.24\,$GeV.  The ultraviolet cutoff, $\Lambda_{\rm uv}$, and running coupling, $\alpha_{\rm IR}$ are chosen in tandem so as to ensure a good description of pseudoscalar meson masses and leptonic decay constants, with the results listed here.  $m$ is the current-mass of the identified quark and $M$ is the associated dressed-quark mass obtained by solving the gap equation.
Empirically, at a sensible level of precision \cite{Zyla:2020zbs}:
$m_\pi =0.14$, $f_\pi=0.092$; $m_K=0.50$, $f_K=0.11$; $m_{\eta_c} =2.98$, $f_{\eta_c}=0.24$; $m_{\eta_b}=9.40$.
%
(Dimensioned quantities in GeV.)}
\begin{center}
\begin{tabular*}
{\hsize}
{
c@{\extracolsep{0ptplus1fil}}|
c@{\extracolsep{0ptplus1fil}}
|c@{\extracolsep{0ptplus1fil}}
c@{\extracolsep{0ptplus1fil}}
|c@{\extracolsep{0ptplus1fil}}
c@{\extracolsep{0ptplus1fil}}
c@{\extracolsep{0ptplus1fil}}}\hline\hline
quark & $\alpha_{\rm IR}/\pi\ $ & $\Lambda_{\rm uv}$ & $m$ &   $M$ &  $m_{0^-}$ & $f_{0^-}$ \\\hline
$l=u/d\ $  & $0.36\phantom{1}$ & $0.91\ $ & $0.007\ $ & 0.37$\ $ & 0.14 & 0.10  \\\hline
$s$  & $0.36\phantom{1}$ & $0.91\ $ & $0.17\phantom{7}\ $ & 0.53$\ $ & 0.50 & 0.11 \\\hline
$c$  & $0.053$ & $1.89\ $ & $1.23\phantom{7}\ $ & 1.60$\ $ & 2.98 & 0.24 \\\hline
$b$  & $0.012$ & $3.54\ $ & $4.66\phantom{7}\ $ & 4.83$\ $ & 9.40 & 0.41
\\\hline\hline
\end{tabular*}
\end{center}
\end{table}

In this subsection, we exploit the SCI detailed in Ref.\,\cite[Appendix~A]{Yin:2021uom}.  Namely, the quark+antiquark scattering kernel is
\begin{align}
{\mathpzc K}_{tu}^{rs}(k,l;P)
& =\frac{4\pi \alpha_{\rm IR}}{m_G^2}  [i\gamma_\mu]_{ts} [i\gamma_\mu]_{ru}\,,
\end{align}
with $m_G=0.5\,$GeV and $\alpha_{\rm IR}$ running with current-quark mass, as listed in Table~\ref{Tab:DressedQuarks}.

Using the SCI, the dressed-quark propagator acquires the following simple form:
\begin{equation}
\label{EqSk}
S(k) = 1/[i\gamma\cdot k + M]\,,
\end{equation}
where $M$ is the dressed-quark mass, which is momentum-independent in this case.  This mass is obtained by solving the SCI gap equation; and the process delivers the values listed in Table~\ref{Tab:DressedQuarks} \cite{Yin:2021uom}.

Similarly, using the SCI, all mesons are described by a Bethe-Salpeter amplitude that is also independent of relative momentum; and for vector-mesons, this amplitude has the form:\footnote{In QCD, the vector-meson Bethe-Salpeter amplitude has eight distinct tensor structures \cite{LlewellynSmith:1969az, Nakanishi:1969ph}, each one multiplied by a different scalar function of both relative and total momentum.}
\begin{align}
\label{EqEV}
\Gamma^\lambda_{V} &= \gamma\cdot \epsilon^\lambda(Q)\, E_{V}\,,
\end{align}
with $E_V$ a constant.  In being described by a bound-state amplitude that is independent of relative momentum, the SCI meson is pointlike in many respects.

The associated meson mass is obtained by solving the SCI Bethe-Salpeter equation in the vector channel, which is simply a one-line algebraic equation \cite{Roberts:2011wy}:
\begin{equation}
0 = 1 + {\cal K}_V (Q^2)\,,
\end{equation}
where
\begin{align}
{\cal K}_V (Q^2) &= -\frac{4\alpha_{\rm IR}}{3\pi m_G^2} \nonumber \\
&\times
\int_0^1d\alpha\, \alpha(1-\alpha)\,Q^2\, \overline{\mathpzc C}_1(\omega(M,\alpha,Q^2))\,,
\label{EqKV}
\end{align}
in which $\overline{\mathpzc C}_1(\omega) = \Gamma(0,\omega \tau_{\rm ir}^2) - \Gamma(0,\omega \tau_{\rm uv}^2)$, with $\Gamma(\alpha,y)$ being the incomplete gamma-function; $\omega(M,\alpha,Q^2)=M^2+\alpha(1-\alpha)Q^2$; and $M$ is the dressed mass of the current-quark that defines the vector-meson.

The value of $E_{V}$, the momentum-independent strength of the Bethe-Salpeter amplitude in Eq.\,\eqref{EqEV}, is fixed by the canonical normalisation condition \cite{LlewellynSmith:1969az, Nakanishi:1969ph, Roberts:2010rn, Roberts:2011wy}:
\begin{equation}
\frac{1}{E_V^2} = -\frac{9 m_G^2}{4\pi \alpha_{\rm IR}}
\frac{d}{dz}\left. {\cal K}_V (z)\right|_{z=-m_V^2}.
\end{equation}

In terms of the canonically normalised amplitude, using Eq.\,\eqref{EqfVmV}, one finds:
\begin{align}
f_V m_V & = \frac{9 m_G^2}{8\pi\alpha_{\rm IR}} E_V\,.
\end{align}

Using the parameters given in Table~\ref{Tab:DressedQuarks}, the following results are obtained \cite{Yin:2021uom}:
\begin{equation}
\label{EqResults}
\begin{array}{c|ccc}
V & E_V & m_V/{\rm GeV} & f_V/{\rm GeV} \\\hline
\rho & 1.53 & 0.93 & 0.13 \\
\phi & 1.63 & 1.03 & 0.12 \\
J/\psi & 1.19 & 3.19 & 0.20 \\
\Upsilon & 1.48 & 9.49 & 0.38
\end{array}\,.
\end{equation}

The inhomogeneous Bethe-Salpeter equation for the dressed photon-quark vertex, Eq.\,\eqref{EqBS}, and its solution, also take simple forms when using the SCI \cite{Roberts:2010rn, Roberts:2011wy}:
\begin{subequations}
\begin{align}
\Gamma_\nu^\gamma(k;Q) & = P_{\rm T}(Q^2) \, T_{\nu\sigma}(Q)  \gamma_\sigma + \gamma_\nu \gamma\cdot Q/Q^2 \,,\\
P_{\rm T}(Q^2) & = \frac{1}{1+{\cal K}_V(Q^2)}\,.
\end{align}
\end{subequations}
Using Eq.\,\eqref{EqKV}, it is clear that $P_{\rm T}(Q^2=0) =1$; consequently, with the aid of Eq.\,\eqref{EqSk}, one readily finds that the Ward-Green-Takahashi identities in Eqs.\,\eqref{SWGTI} are also preserved; and, similarly, Eqs.\,\eqref{EqPiWGTI}\,--\,\eqref{QEDVP}, so that the photon remains massless.

The question of the fidelity of the VMD assumption is now readily posed and addressed within the SCI framework.  Namely, expressing Eq.\,\eqref{EqVMDvertex}, it is apparent from the formulae written above that
\begin{align}
& \epsilon^\lambda \cdot \Gamma^\gamma(Q) \stackrel{Q^2+m_V^2\simeq 0}{=}
\epsilon^\lambda\cdot\gamma \frac{2 f_V m_V E_V }{Q^2+m_V^2} \,;
\end{align}
to wit, one recovers Eq.\,\eqref{EqCFI} on the meson mass shell.  (This identity is readily verified numerically.)  Hence, the reliability of the VMD assumption may be measured by the deviation of the following ratio from unity:
\begin{equation}
\left. R_V(Q^2) \right|_{Q^2=0} = \left. \frac{1}{P_{\rm T}(Q^2)}\frac{2 f_V m_V E_V }{Q^2+m_V^2 }\right|_{Q^2=0} = \frac{2 f_V E_V}{m_V}\,.
\label{RatioVMD}
\end{equation}
Using the results in Eq.\,\eqref{EqResults}, one finds
\begin{equation}
\begin{array}{l|cccc}
V & \rho & \phi & J/\psi & \Upsilon \\\hline
R_V(0) & 0.42 & 0.37 & 0.15 & 0.12\\
\end{array}\,.
\end{equation}
Evidently, for vector-mesons composed of the lighter quarks, use of the VMD assumption leads one to overestimate the connection in cross-sections between $e + p \to e^\prime + V + p$ and $V + p \to V+p$ by a factor of six and this overestimate exceeds a factor of fifty for vector-mesons composed of heavy quarks.  (Recall that a cross-section is obtained from the amplitude-squared.)  Since the SCI produces photons and vector-mesons that both possess a pointlike character, these poor outcomes are likely the best achievable, so far as the fidelity of the  VMD assumption is concerned in connection with vector-meson photoproduction.  Naturally, the results are worse for electroproduction ($Q^2>0$).

Notwithstanding these things, the pointlike features of SCI bound-states ensure that the mismatch between $e + p \to e^\prime + V + p$ and $V + p \to V+p$ is merely an overall $Q^2$-dependent multiplicative term whose impacts may be removed by including an off-shell form factor \cite{Cao:2019kst, Wu:2019adv}:
\begin{equation}
\label{gammaVFQ2}
\gamma_{\gamma V} \to F_{\gamma V}(Q^2) \gamma_{\gamma V} \,.
\end{equation}
We now turn to the QCD-like cases, wherein all quantities depend on both relative and total momentum.

%

\subsection{Momentum-dependent interaction}
\label{SecReal}
In connection with RL truncation, a realistic momentum-dependent interaction was introduced in Refs.\,\cite{Qin:2011dd, Qin:2011xq}:
\begin{align}
\label{defcalG}
 \tfrac{1}{Z_2^2}\tilde{\mathpzc G}(s) & =
 \frac{8\pi^2}{\omega^4} D e^{-s/\omega^2} + \frac{8\pi^2 \gamma_m \mathcal{F}(s)}{\ln\big[ \tau+(1+s/\Lambda_{\rm QCD}^2)^2 \big]}\,,
\end{align}
where $\gamma_m=12/25$, $\Lambda_{\rm QCD}=0.234\,$GeV, $\tau={\rm e}^2-1$, and ${\cal F}(s) = \{1 - \exp(-s/[4 m_t^2])\}/s$, $m_t=0.5\,$GeV.  This form has been used widely; and amongst the more recent applications, predictions for vector-meson elastic form factors \cite{Xu:2019ilh} and semileptonic $B_c\to \eta_c, J/\psi$ transitions \cite{Yao:2021pyf} are most closely related to the analysis herein.
As explained, \emph{e.g}., in Ref.\,\cite[Sec.\,II.B]{Xu:2019ilh}: $D\omega = (0.8\,{\rm GeV})^3$, $\omega = 0.5\,$GeV for $u=d$-, $s$-quarks; and $D\omega = (0.6\,{\rm GeV})^3$, $\omega = 0.8\,$GeV for $c$-, $b$-quarks.  The shift in parameter values owes to the diminishing strength of corrections to RL truncation as one moves into the heavy-quark sector.   (See, \emph{e.g}., Ref.\,\cite{Bhagwat:2004hn}.)

Details relating to the development of Eq.\,\eqref{defcalG} and its connection with QCD are presented in Refs.\,\cite{Qin:2011dd, Qin:2011xq, Binosi:2014aea}.  Here, we simply reiterate some points.
(\emph{i}) The interaction is consistent with that found in studies of QCD's gauge sector.  It expresses the result, enabled by strong non-Abelian gauge-sector dynamics, that the gluon propagator is a bounded, smooth function of spacelike momenta, whose maximum value on this domain is at $s=0$ \cite{Binosi:2016xxu, Gao:2017uox, Cui:2019dwv}, and capitalises on the property that the dressed gluon-quark vertex does not possess any structure which can qualitatively alter these features \cite{Kizilersu:2021jen}.
(\emph{ii}) Eq.\,\eqref{defcalG} preserves the one-loop renormalisation group behaviour of QCD; hence, \emph{e.g}., the quark mass-functions produced are independent of the renormalisation point.
(\emph{iii}) On $s \lesssim (2 m_t)^2$, Eq.\,\eqref{defcalG} defines a two-parameter \emph{Ansatz}, the details of which determine whether such corollaries of EHM as confinement and dynamical chiral symmetry breaking (DCSB) are realised in solutions of the bound-state equations \cite{Roberts:2020hiw, Roberts:2021nhw}.
Additionally, given a value of the product $D \omega$, results for observables are practically insensitive to variations $\omega \to \omega (1\pm 0.1)$; thus, there is no issue of fine tuning.
(\emph{iv}) The interaction is specified in Landau gauge because, amongst other things, this gauge is a fixed point of the renormalisation group and minimises sensitivity to the form of the gluon-quark vertex, thus providing the conditions for which RL truncation is most accurate.  
Naturally, all Schwinger functions considered herein are gauge covariant; hence, whilst quantitative characteristics respond to properly implemented changes in gauge, qualitative features and observable quantities are gauge independent.

Employing Eqs.\,\eqref{KDinteraction}, \eqref{defcalG} to complete the gap and Be\-the-Salpeter equations, one obtains the results in Table~\ref{DSErealistic}: the mean absolute relative difference between calculation and experiment is $4.5$\%, with median value 1.5\%.
It is worth highlighting that Table~\ref{DSErealistic} lists renormalisation-point-invariant current-quark masses.  One-loop evolution to $\zeta=\zeta_2=2\,$GeV yields $m_u^{\zeta_2}=0.0046\,$GeV; $m_s^{\zeta_2}=0.112\,$GeV; and solving for the one-loop heavy-quark mass produces $m_c^{m_c}=1.19\,$GeV, \linebreak $m_b^{m_b}=4.38\,$GeV.  All these values are commensurate with those typically quoted \cite{Zyla:2020zbs}.
%

\begin{table}[t]
\caption{\label{DSErealistic}
Masses and decay constants obtained using the bound-state equations defined by Eqs.\,\eqref{KDinteraction}, \eqref{defcalG}: $\hat m$ is the renormalisation-group-invariant current-quark mass for the identified quark; $0^-$ labels ground-state pseudoscalar systems; and $V$, ground-state vector states.  The $0^-_{s\bar s}$ state is a fictitious pseudoscalar meson computed as a benchmark in both continuum and lattice analyses, \emph{e.g}., Ref.\,\cite{Chen:2018rwz}.
For comparison, where known, empirical values are \cite{Zyla:2020zbs}:
$m_\pi=0.138$, $m_{\eta_c} = 2.98$, $m_{\eta_b}=9.40$,
$f_\pi = 0.092$, $f_{\eta_c}=0.24$,
$m_\rho = 0.775$, $m_\phi=1.019$, $m_{J/\psi} = 3.10$, $m_\Upsilon = 9.46$
$f_\rho = 0.156(1)$, $f_\phi = 0.161(3)$, $f_{J/\psi} = 0.29$, $f_\Upsilon = 0.51$.
(Dimensioned quantities in GeV.)}
\begin{center}
\begin{tabular*}
{\hsize}
{
l@{\extracolsep{0ptplus1fil}}|
c@{\extracolsep{0ptplus1fil}}
c@{\extracolsep{0ptplus1fil}}
c@{\extracolsep{0ptplus1fil}}
c@{\extracolsep{0ptplus1fil}}
c@{\extracolsep{0ptplus1fil}}}\hline
quark$\ $ & $\hat m\ $ & $m_{0^-}\ $ & $f_{0^-}\ $ & $m_V\ $ & $f_V\ $ \\\hline
$u=d$ & $0.00664\ $ & $0.138\ $ & $0.093\ $ & $0.735\ $ & $0.146\ $  \\
$s$ & $0.162\phantom{00}\ $ & $0.691\ $ & $0.130\ $ & $1.074\ $ & $0.183\ $ \\
$c$ & $1.51\phantom{000}\  $
    & $2.98\phantom{0}\ $ & $0.284\ $ & $3.12\phantom{0}\ $ & $0.300\ $ \\
$b$ & $7.34\phantom{000}\ $ & $9.26\phantom{0}\ $& $0.475\ $ & $9.33\phantom{0}\ $& $0.505\ $\\\hline
\end{tabular*}
\end{center}
\end{table}

We are now in a position to test Eq.\,\eqref{RatioVMD} in the realistic case of nonpointlike vector-mesons produced by a QCD-constrained momentum-dependent interaction.  The necessary generalisation brings some complications because, as noted above, the Poincar\'e-covariant Be\-the-Sal\-peter amplitude associated with a nonpointlike vector-meson has eight independent components, each one multiplied by an associated Poincar\'e-invariant scalar function expressing a momentum-dependent strength factor \cite{LlewellynSmith:1969az}.  (Illustrative numerical solutions are drawn elsewhere \cite[Sec.\,V]{Maris:1999nt}.)  The dominant amplitude is that associated with $\gamma\cdot \epsilon(Q)$, which is conventionally written as $F_1(k^2,k\cdot Q; Q^2)$: the other seven functions in $\Gamma_V^\lambda(k;Q)$ are driven to be nonzero by $F_1(k^2,k\cdot Q; Q^2) \neq 0$.  Consequently, it is sufficient to work with $\gamma\cdot \epsilon^\lambda(Q)$ $\times F_1(k^2,k\cdot Q; Q^2)$.  For subsequent use, we note that the zeroth Chebyshev moment of this or any analogous function is ($k\cdot Q =: x \sqrt{k^2 Q^2} $):
\begin{equation}
F_1^0(k^2;Q^2) = \frac{2}{\pi}\int_{-1}^1 dx\,\sqrt{1-x^2}\,F_1(k^2,k\cdot Q; Q^2)\,.
\end{equation}

Regarding the other side of the equation, the dressed photon-quark vertex has twelve independent structures, eight of which are essentially transverse \cite{Ball:1980ay, Curtis:1990zs}, contributing nothing to resolving the Ward-Green-Takaha\-shi identity and expressing all on-shell overlap with any vector-meson bound-states.  Indeed, on the mass-shell of any vector-meson, just as with the contact interaction:
\begin{align}
 \epsilon^\lambda \cdot \Gamma^\gamma(k;Q) & \stackrel{Q^2+m_V^2\simeq 0}{=}
 \frac{2 f_V m_V }{Q^2+m_V^2} \Gamma_V^\lambda(k;Q) \,.
 \label{RatioReal}
\end{align}
In the photon-quark vertex, the leading component is also that associated with $\gamma\cdot \epsilon^\lambda(Q)$.

The VMD issue we are exploring was studied in Ref.\,\cite{Maris:1999bh}, with a focus on the $\rho$-meson.  Therein, owing to the Ward identity, Eq.\,\eqref{Ward}, it was concluded that there is zero overlap between a real massless photon and a massive composite vector-meson at $Q^2=0$, \emph{i.e}., $\epsilon^\lambda \cdot \Gamma^\gamma(k;Q)|_{Q^2\simeq 0}$ receives no contribution from any such vector-meson bound-state. Mathematically, this is true, as we demonstrated in Sec.\,\ref{SecPVP}.  However, given the phenomenological character of the VMD \emph{Ansatz}, we choose to admit the possibility of a serviceable correspondence between $\epsilon^\lambda \cdot \Gamma^\gamma|_{Q^2\simeq 0}$ and $\Gamma_V^\lambda(k;Q)|_{Q^2+m_V^2\simeq 0}$ and seek evidence for or against this hypothesis.

In this case, with momentum-dependent interactions and vertices, we begin to address the issue of the fidelity of the VMD assumption by comparing the zeroth Chebyshev moments of the functions multiplying the leading matrix structure on both sides of Eq.\,\eqref{RatioReal}.\footnote{All other moments and functions are more sensitive to vector-meson compositeness and nonlocality; so if the fidelity is poor for the zeroth moment of the dominant amplitude, it will be worse for every other possible comparison.}  These functions are readily obtained using a suitably chosen projection operator; and defining this leading term in $\epsilon^\lambda(Q)\cdot \Gamma^\gamma(k;Q)$ to be $\gamma\cdot \epsilon^\lambda(Q) G_1(k^2,k\cdot Q; Q^2)$, the following comparison arises for consideration:
\begin{align}
G_1^0(k^2; Q^2=0) \; \mbox{vs.}\; \frac{2 f_V}{m_V} F_1^0(k^2;-m_V^2)\,.
\end{align}
Namely, one measure of the accuracy of the VMD assumption is the ratio:
\begin{equation}
\label{RVk2Q20}
R_V(k^2;Q^2=0):= \frac{2 f_V}{m_V} \frac{F_1^0(k^2;-m_V^2)}{G_1^0(k^2; Q^2=0)}\,,
\end{equation}
which is the obvious analogue of Eq.\,\eqref{RatioVMD}.  As here, it is momentum-dependent in QCD; and only if $R_V(k^2;Q^2=0) \approx 1$ can the VMD assumption be considered reliable.  Naturally, referring to Eq.\,\eqref{GenS}, and using the Ward identity, Eq.\,\eqref{Ward},
\begin{equation}
G_1^0(k^2; Q^2=0) = A(k^2)\,,
\end{equation}
where the appropriate form for $A(k^2)$ is obtained from the gap equation for the relevant quark flavour.

\begin{figure}[t] 
\vspace*{2ex}

\leftline{\hspace*{0.5em}{\large{\textsf{A}}}}
\vspace*{-3ex}
\includegraphics[width=0.42\textwidth]{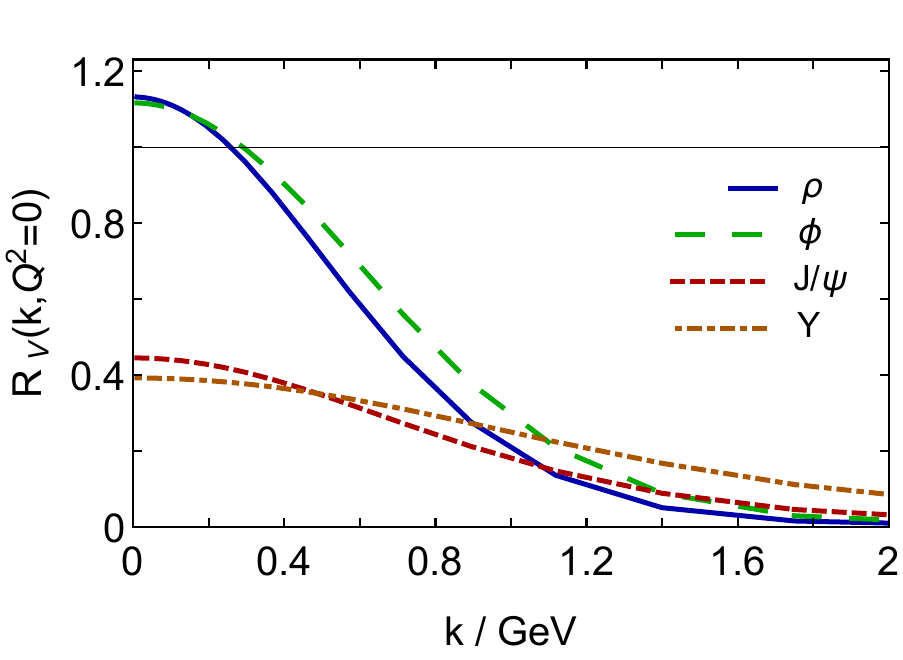}
\vspace*{-1ex}

\leftline{\hspace*{0.5em}{\large{\textsf{B}}}}
\vspace*{-3ex}
\includegraphics[width=0.42\textwidth]{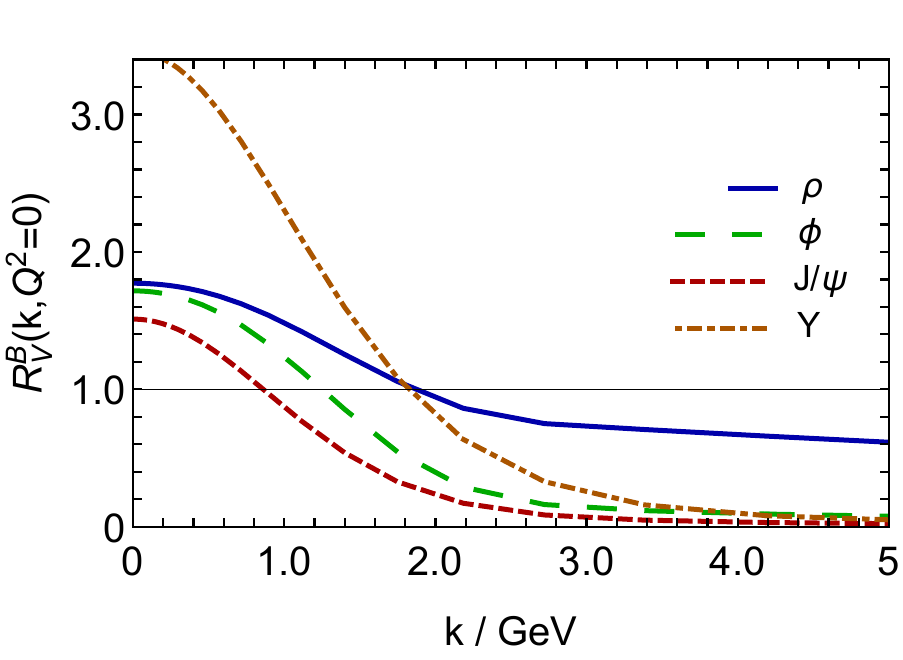}
\caption{\label{FigRVk2Q20}
\emph{Upper panel}\,--\,{\sf A}: Ratio in Eq.\,\eqref{RVk2Q20} computed using matched solutions of the gap and Bethe-Salpeter equations defined by Eqs.\,\eqref{KDinteraction}, \eqref{defcalG} for $V=\rho, \phi, J/\psi, \Upsilon$.
\emph{Lower panel}\,--\,{\sf B}: Ratio in Eq.\,\eqref{RVk2Q20B}, computed analogously.
In cases where the VMD hypothesis were sound, all these curves would lie close to the thin horizontal line drawn at unity.
}
\end{figure}

We computed the dimensionless ratio in Eq.\,\eqref{RVk2Q20} for each vector-meson in Table~\ref{DSErealistic} and depict the results in Fig.\,\ref{FigRVk2Q20}A.  As expected mathematically, since the on-shell photon is a massless pointlike object and an on-shell vector-meson is a massive composite object, the ratio $R_V(k^2;Q^2=0)$ reveals that there is no overlap between these two states at $Q^2=0$.  In QCD, vector-meson compositeness requires that $F_1^0(k^2;-m_V^2)$ vanish as $1/k^2$ with increasing $k^2$, up to logarithmic corrections, whereas for the pointlike photon, $G_1^0(k^2; Q^2=0) \to 1$ with increasing $k^2$, again up to logarithmic corrections.

To complete the picture, it is worth analysing a different term in the photon-quark vertex, \emph{viz}.\ the piece whose appearance in the absence of Higgs couplings into QCD is a direct consequence of DCSB, which is a corollary of EHM.  Once again, this term can be read using the Ward identity, Eq.\,\eqref{Ward}:
\begin{equation}
G_3(k^2;Q^2=0) = -2\frac{d}{dk^2} B(k^2)\,.
\end{equation}
Denoting the related term in the vector-meson Bethe-Salpeter amplitude by $F_8(k^2,k\cdot Q;-m_V^2)$, one is led to consider the ratio:
\begin{equation}
\label{RVk2Q20B}
R_V^B(k^2;Q^2=0):= \frac{f_V}{m_V} \frac{F_8^0(k^2;-m_V^2)}{G_3^0(k^2; Q^2=0)}\,,
\end{equation}
which is plotted for each vector-meson in Fig.\,\ref{FigRVk2Q20}B.  The observations made in connection with Fig.\,\ref{FigRVk2Q20}A are equally applicable here, except for the fact that both the numerator and denominator here fall as $1/k^2$ up to logarithmic corrections.

Comparing the curves in both panels of Fig.\,\ref{FigRVk2Q20}, within each panel and across panels, it becomes apparent that the momentum-dependence in any quark+anti-quark loop describing scattering in the process $e + p \to e^\prime + V + p$ is very different from that in the process $V+p \to V+p$.  Moreover, this mismatch is compounded by the fact that the $Q^2=0$ photon-quark vertex has only three nontrivial momentum-dependent components, whereas the on-shell vector-meson has eight distinct, momentum-dependent terms.  Hence, only one other $Q^2=0$ ratio is nonzero.  The other five are identically zero.
Consequently, an expedient like that in Eq.\,\eqref{gammaVFQ2} is inadequate in the realistic case.

%
%

It is important to reiterate that analyses based on Eqs.\,\eqref{KDinteraction}, \eqref{defcalG} have delivered sound results for the properties of ground-state vector-mesons, and also many other systems, including baryons \cite{Qin:2018dqp, Wang:2018kto, Qin:2019hgk}.  This being so, then it is justified to draw conclusions about physical processes from the material in this subsection.  Hence, in light of its revelations, the best hope for a measure of usefulness in the VMD assumption is that damping introduced by the wave functions of the initial- and final-state protons and the final-state vector-meson restricts the loop integration(s) that express the process $q + \bar q + p \to V + p$ to a domain that is not overwhelmingly sensitive to the differences between the ``wave functions'' of a pointlike as compared to a composite object; namely, to a domain throughout which $R_V(k^2;Q^2=0) $ and its nonzero analogues are $\approx 1$ and the other nonzero terms in the initial-state vector-meson Bethe-Salpeter amplitude may be neglected.

Regarding Fig.\,\ref{FigRVk2Q20}, this might be plausible for the $\rho$-meson and, with less confidence, the $\phi$-meson, so that the VMD assumption may be useable in these channels in some circumstances.  However, again, one must bear in mind that five of the structures in the vector-meson Bethe-Salpeter amplitude are ignored by the VMD assumption.

On the other hand, confirming the result obtained by other means in Ref.\,\cite{Xu:2019ilh}, it is not the case for the heavy mesons.  So, it is unsound to use $e^+ e^-$ decays of these systems in order to estimate effective strength factors, $\gamma_{\gamma J/\psi}$, $\gamma_{\gamma \Upsilon}$.
The force of this conclusion is magnified by the following facts:
$R_V(k^2;0)$ in Fig.\,\ref{FigRVk2Q20}A compares only the zeroth Chebyshev moments of the leading amplitudes;
the situation for all other moments and amplitudes is significantly worse, as highlighted by \linebreak Fig.\,\ref{FigRVk2Q20}B and especially because the VMD \emph{Ansatz} omits five components of the vector-meson Bethe-Salpe\-ter amplitude;
and these last remarks emphasise once more that no single overall multiplicative function can remedy the diverse array of mismatched momentum dependences.

Consequently, the VMD assumption is false for heavy vector-mesons.  Furthermore, there is no model-indepen-dent way to estimate and/or correct for the degree by which it distorts any interpretation of the $e + p \to e^\prime + V + p$ reaction in terms of the $V+p \to V+p$ process; and it is only the latter for which a QCD multipole expansion has been used to draw connections with the proton's glue distribution \cite{Krein:2020yor}.
Stated figuratively, in order to ensure a quantitatively accurate analysis of $e + p \to e^\prime + V + p$, one would need to use
\begin{equation}
\gamma_{\gamma V} \to \gamma_{\gamma V} {\mathpzc M}(k^2,k\cdot Q,Q^2)\,,
\end{equation}
where ${\mathpzc M}(k^2,k\cdot Q,Q^2)$ is a matrix-valued function whose reliable estimation will only become possible after development of a realistic reaction theory for $q + \bar q + p \to V + p$.

It is worth remarking that these conclusions are consistent with dispersion theory.  For example, consider the usual spectral representation of the elastic electromagnetic form factor of a light-quark hadron.  In this case, the spectral function, $\sigma(t)$, has a prominent feature associated with the $\rho$-meson, broadened by its hadronic width, not too far removed from the $\pi^+ \pi^-$ production threshold.  The hadron's electromagnetic radii receive a modest contribution from this part of $\sigma(t)$; but the exact amounts depend on how one chooses to draw the boundaries on this $\rho$-meson spectral feature \cite[Sec.\,2.3]{Roberts:2000aa}.   For analogous cases involving heavy vector-mesons, the kindred spectral feature lies deep in the timelike region, \emph{e.g}., $t=16\,m_\rho^2$ for the $J/\psi$ and $t=150\,m_\rho^2$ for the $\Upsilon$, and is narrow in both instances.  Here, the spectral strength in the neighbourhood $t\simeq 0$ is not dominated by those distant bound-state features, but, instead, by nonresonant quark+antiquark scattering processes.

One may compare these conditions with those prevailing when using Sullivan-like processes \cite{Sullivan:1971kd} to explore the structure of $\pi$ and $K$ targets, \emph{e.g}., $e + p \to e^\prime + \pi(K) + n$ or $e + p \to e^\prime + X + n$.  In such cases, the off-shell quark+antiquark correlation serves as a valid $\pi(K)$ target on $-t \lesssim 1 (1.5)\,m_\rho^2$ \cite{Qin:2017lcd}, which is the domain to which completed and planned experiments restrict themselves \cite{Aguilar:2019teb, Volmer:2000ek, Horn:2006tm, Tadevosyan:2007yd, Blok:2008jy, Huber:2008id, Horn:2007ug, E12-19-006, Carmignotto:2018uqj, E12-09-011, Arrington:2021biu, Roberts:2021nhw}.

\section{Summary and Perspective}
\label{epilogue}
We examined the fidelity of the single-pole vector meson dominance (VMD) hypothesis as a tool for interpreting vector-meson photo/electroproduction reactions, like $e + p \to e^\prime + V + p$, as an access route to a different, desired reaction; to wit, $V + p \to V+p$ in the exemplifying case.

As the first step in this study, we considered the photon vacuum polarisation tensor, $\Pi_{\mu\nu}(Q)$, where $Q$ is the photon momentum; and reaffirmed that there is no vector-meson contribution to this polarisation at $Q^2=0$, \emph{i.e}., the photoproduction point [Sec.\,\ref{SecPVP}].  This outcome reveals that massless real photons are readily distinguishable from massive vector-bosons and the current-field identity [Eq.\,\eqref{EqCFI}], typical of VMD implementations, cannot be used literally and alone because it leads to violations of Ward-Green-Takahashi identities in quantum electrodynamics.

We then considered the dressed photon-quark vertex, $\Gamma_\nu^\gamma(k;Q)$.  It possesses a pole at the mass of any vector-meson bound-state, which is a physical property, expressing the fact that $V\to e^+ e^-$ decay proceeds via a timelike virtual photon.  From this perspective, the VMD hypothesis may be viewed as an assertion that $\left.\Gamma_\nu^\gamma(k;Q)\right|_{Q^2\simeq 0}$ maintains an unambiguous link in both strength and momentum-dependence with the bound-state amplitude of an on-shell vector-meson [Sec.\,\ref{SecGQvtx}].

We explored this possibility using two models for the quark+quark scattering kernel.  Supposing that the kernel is momentum-independent [Sec.\,\ref{SecCI}], then such a link does exist because the vector-mesons produced by a contact interaction are described by momentum-independent bound-state amplitudes.  On the other hand, any use of the $V\to e^+ e^-$ decay width to estimate the coupling strength via the current-field identity leads one to overestimate the relation between the cross-section for $e + p \to e^\prime + V + p$ and that for $V + p \to V+p$ by a factor of $\sim 50$  for heavy vector-mesons.

In the realistic case [Sec.\,\ref{SecReal}], where the quark+quark scattering kernel is momentum dependent, as it is in QCD, the VMD hypothesis is false for heavy-mesons because the momentum-de\-pen\-dence of the $Q^2=0$ photon-quark vertex is entirely different from that of the vector-meson Bethe-Salpeter amplitude.  Hence, the process $\gamma^{(\ast)}(Q^2\geq 0) + p \to V + p$ has no discernible link to $V+p \to V+p$, either in strength or in the momentum dependence of the integrands that would appear in computing the different reaction amplitudes.

Given that the momentum-dependent kernel we used to complete the analysis herein is known to provide a realistic description of ground-state vector-mesons and many other systems, including ground-state baryons, this conclusion may reasonably be transferred to QCD processes.  If that is so, then no existing attempt to connect $e + p \to e^\prime + V + p$ reactions with $V + p \to V+p$ via VMD can be viewed as reliable.  Amongst other things, this makes tenuous any interpretation of $e + p \to e^\prime + V + p$ reactions as a route to hidden-charm pentaquark production or as a means of uncovering the origin of the proton mass.

As demonstrated in numerous applications \cite{Eichmann:2016yit, Roberts:2020udq, Eichmann:2020oqt, Roberts:2020hiw, Qin:2020rad}, including $\gamma^\ast \gamma \to \pi^0, \eta, \eta^\prime, \eta_c, \eta_b$ \cite{Raya:2016yuj, Ding:2018xwy}, a viable alternative to the VMD hypothesis consists in adapting the continuum Schwinger function methods we have employed herein to directly analyse processes like $\gamma^{(\ast)} + p \to V + p$.  Regarding vector-meson photo/electropro-duction from the proton, Ref.\,\cite{Pichowsky:1996tn} illustrates how one might proceed.  Given developments in the past vicennium, it is now possible to improve upon such studies.

\begin{acknowledgements}
%
%
We are grateful for constructive comments from G.~Krein, T.-S.\,H.~Lee and P.-L.~Yin.
Work supported by:
Nanjing University Innovation Programme for PhD candidates;
National Natural Science Foundation of China (under Grant No.\,11805097);
and
Jiangsu Provincial Natural Science Foundation of China (under Grant No.\,BK20180323).
\end{acknowledgements}


\end{document}